\documentclass[%
 reprint,
superscriptaddress,
 amsmath,amssymb,
 aps,
]{revtex4-2}

\usepackage{graphicx}% Include figure files
\usepackage{dcolumn}% Align table columns on decimal point
\usepackage{bm}% bold math
\usepackage{braket}
\usepackage{caption}
\usepackage{subcaption}
\usepackage{color}
\begin{document}

%\preprint{APS/123-QED}

\title{Coupled cluster method tailored by quantum selected configuration interaction}% Force line breaks with \\

\newcommand{\affilA}{%
Center for Quantum Information and Quantum Biology,
The University of Osaka, 1-2 Machikaneyama, Toyonaka, Osaka 560-0043, Japan
}%

\author{Luca~Erhart}
\affiliation{\affilA}
\author{Yuichiro~Yoshida}
\email{yoshida.yuichiro.qiqb@osaka-u.ac.jp}
\affiliation{\affilA}
\author{Wataru~Mizukami}
\affiliation{\affilA}

% \altaffiliation[Also at ]{Physics Department, XYZ University.}%Lines break automatically or can be forced with \\
%\author{Second Author}%
% \email{Second.Author@institution.edu}
%\affiliation{%
% Authors' institution and/or address\\
% This line break forced with \textbackslash\textbackslash
%}%

\date{\today}% It is always \today, today,
             %  but any date may be explicitly specified

\begin{abstract}

We present the quantum-selected configuration interaction-tailored coupled-cluster (QSCI-TCC) method, a hybrid quantum-classical scheme that tailors coupled-cluster (CC) theory with a quantum-selected configuration interaction (QSCI) wave function.
QSCI provides a scalable, shot-efficient approach to reconstructing the many-electron state prepared on quantum hardware on a classical computer.
The resulting active-space CI coefficients, which are free from additive shot noise, are mapped to fixed cluster amplitudes within the tailored coupled-cluster framework, 
after which a conventional CC calculation optimizes the remaining amplitudes.
This workflow embeds static (strong) correlation from the quantum device and subsequently recovers dynamical (weak) correlation, yielding a balanced description of both.
The method is classically simulated and applied to the simultaneous O--H bond dissociation in H$_2$O and the triple-bond dissociation in N$_2$.
QSCI-TCC and its perturbative-triples variant, QSCI-TCC(T), provide accurate results even where CCSD or CCSD(T) begin to break down.
Shot-count tests for the N$_2$ (6e, 6o) active space demonstrate that, with the (c) correction, chemically sufficient precision ($\leq 1$~kcal/mol) is achieved with only $1.0 \times 10^5$ shots in the strongly correlated regime ($r=2.2$~\AA)---an order of magnitude fewer than required by an earlier matchgate-shadows implementation [J. Chem. Theory Comput., {\bf 20}, 5068 (2024)].
By pairing resource-efficient quantum sampling with the CC theory, QSCI-TCC provides a promising pathway to quantum-chemical calculations of classically intractable systems.

\end{abstract}

%\keywords{Suggested keywords}%Use showkeys class option if keyword
                              %display desired
\maketitle

%\tableofcontents

%\section{\label{sec:level1}First-level heading:\protect\\ The line
%break was forced \lowercase{via} \textbackslash\textbackslash}

\section{Introduction \label{sec:intro}}

Quantum computers have advanced rapidly, and quantum chemistry is now regarded as one of the most promising applications of quantum computing.
The considerable attention to this field originates from the fact that the quantum phase estimation (QPE) algorithm~\cite{Abrams1997simulation,Abrams1999quantum} can determine eigenstates of the electronic structure Hamiltonian in polynomial time, whereas the best-known classical algorithms scale exponentially~\cite{AspuruGuzik2005simulated}.
Consequently, reducing the resource cost of fault-tolerant quantum computation has become a central goal of the field~\cite{Reiher2017elucidating,Berry2019qubitization,vonBurg2021quantum,Lee2021even,Dario2024reducing,Guang2025fast}.
Ongoing progress in hardware and algorithms promises to extend quantum-chemical accuracy to a wide range of problems in chemistry and materials science. 

In contrast to the long-term, fault-tolerant scenario outlined above, recent demonstrations of quantum-chemical calculations using current quantum hardware have steadily approached the accuracy and accessible system sizes of advanced conventional electronic-structure methods.
These advances have been enabled by hybrid approaches that embed quantum computations---capturing static (strong) correlation through coherent superpositions---within classical post-Hartree-Fock procedures that recover dynamical (weak) correlation, thus enabling quantitative accuracy.
The integration of quantum computation with auxiliary-field quantum Monte Carlo (AFQMC)~\cite{Huggins2022unbiasing,Yoshida2025auxiliary,Danilov2025enhancing} and coupled cluster (CC)~\cite{Erhart2024,Scheurer2024} exemplifies this strategy.

CC theory is one of the most successful methods in modern quantum chemistry~\cite{Bartlett2007coupled,Bartlett2024perspective}. 
Widely implemented in quantum chemistry software, CC excels at describing dynamical correlation and typically yields high accuracy for single-reference (SR) systems---those lacking significant orbital quasi-degeneracy.
Its polynomial computational cost is attractive: CC singles and doubles (CCSD) scales as $O(N^6)$, while the perturbative triples extension, CCSD(T), scales as $O(N^7)$. 
Further developments, such as linear-scaling CC methods~\cite{Riplinger2013efficient,Riplinger2013natural,Riplinger2016sparse,Saitow2017new,Guo2018communication,Liakos2020comprehensive,Guo2020linear}, have extended CC to relatively small proteins. In addition, CC data now serve as valuable training data for machine learning~\cite{Wilkins2019accurate,Smith2019approaching}.
Therefore, CC theory is significant for SR systems and has widespread applicability.

However, conventional CC cannot describe multi-reference (MR) systems, where quasi-degenerate orbitals yield strong correlation. 
The tailored CC (TCC) method~\cite{Kinoshita2005,Hino2006tailored} addresses this by embedding amplitudes obtained from an MR wave function into the CC ansatz.
While early implementations employed complete active space configuration interaction (CASCI)~\cite{Kinoshita2005,Hino2006tailored}, recent studies have incorporated other MR solvers, such as density matrix renormalization group~\cite{Faulstich2019numerical,Visnak2024DMRG} and full configuration interaction quantum Monte Carlo~\cite{Vitale2020FCIQMC,Vitale2022FCIQMC}, achieving a balanced treatment of strong and weak correlations.

The integration of quantum computation into TCC has recently been proposed~\cite{Erhart2024,Scheurer2024}.
This provides a significant advantage because it avoids estimating higher-order reduced density matrices (RDMs), which are expensive to obtain on a quantum device~\cite{Nishio2023statistical}. 
By contrast, the quantum state realized on a quantum computer is reconstructed via state-tomography techniques and converted directly into CC amplitudes. 
Two initial studies employed computational basis tomography~\cite{Erhart2024} and matchgate shadows tomography~\cite{Scheurer2024},
demonstrating balanced accuracy but highlighting the need for efficient sampling from quantum states.

Quantum-selected configuration interaction (QSCI)~\cite{Kanno2023quantum} is a scalable state tomography protocol that can work with relatively small shot budgets and is highly flexible. It can be combined with various quantum algorithms, such as the variational quantum eigensolver (VQE)~\cite{Peruzzo2014variational} and QPE, to determine the eigenstate of the electronic structure Hamiltonian. 
QSCI has expanded quantum-classical hybrid calculations to 77 qubits~\cite{Robledo2024chemistry}, a scale that already enables practical quantum-chemical applications. 
Note that QSCI has been sometimes referred to as sampling-based quantum diagonalization (SQD)~\cite{Robledo2024chemistry}; however, this study treats them as essentially identical.
In QSCI, an exact quantum state is not always necessary, which reduces some workloads of quantum devices~\cite{Kanno2023quantum}. Furthermore, the hardness of classical sampling from unitary cluster Jastrow circuits suggests a potential quantum advantage in QSCI~\cite{Hafid2025hardness}.

Moreover, QSCI offers a significant advantage: its configuration-interaction (CI) coefficients are free from the additive shot noise that plagues many other tomography schemes. 
In QSCI, computational-basis states are first sampled from the prepared quantum state; although this sampling is stochastic, it affects only which determinants enter the subspace.
The subsequent steps---constructing the effective Hamiltonian within that subspace and diagonalizing it---are performed exactly.
Consequently, the sole error in the CI coefficients stems from subspace truncation, not from statistical noise.
This feature contrasts sharply with computational basis tomography (sampling)~\cite{Kohda2022quantum} and classical shadows methods~\cite{Huang2020predicting,Wan2023matchgate,Heyraud2025unified}, both of which propagate shot noise directly into the reconstructed state.
A noise-free wave function is particularly beneficial for downstream classical algorithms. For example, a quantum-classical hybrid quantum Monte Carlo method was performed stably~\cite{Yoshida2025auxiliary}, as opposed to additive errors that had prevented stable computation. Therefore, it makes QSCI a powerful partner for advanced correlated wave-function methods.

In this study, we introduce a TCC variant tailored with QSCI (denoted as QSCI-TCC).
The QSCI procedure efficiently samples the electronic configurations realized on the quantum device and reconstructs the coherent wave function on a classical computer.
This wave function, which captures the strong correlation, is mapped to CC amplitudes; subsequent CC iterations then recover the remaining dynamical correlation.
We validate QSCI-TCC on several MR systems, such as the simultaneous bond dissociation in H$_2$O and the triple bond dissociation in N$_2$.
We further investigate how finite shot measurements affect their accuracy.

The remainder of this paper is organized as follows. 
Sec~\ref{sec:methods} reviews the TCC and QSCI formalisms and presents an overview of QSCI-TCC. Computational details are given in Sec~\ref{sec:comput_details}. 
Results are discussed in Sec~\ref{sec:results}, and Sec~\ref{sec:conclusions} summarizes our contributions and outlines the current limitations.

  %I believe leaving the sections in separate files is more organized, change it if you desire 
\section{Computational methods \label{sec:methods}}

\subsection{Review of tailored coupled cluster}

TCC~\cite{Kinoshita2005,Hino2006tailored} factorizes the cluster exponential $e^{\hat{T}}$ as
\begin{equation} \label{equ_TCC_tailoring}
    \ket{\Psi_{\text{TCC}}} = e^{\hat{T}} \ket{\Psi_0} 
    = e^{\hat{T}^\text{{rest}}} e^{\hat{T}^{\text{active}}} \ket{\Psi_0},
\end{equation}
where $\ket{\Psi_0}$ is a reference Slater determinant.
The cluster operator is partitioned into an active-space component $\hat{T}^\textrm{active}$, which acts entirely within the active space, and a remainder $\hat{T}^\textrm{rest}$. 
The splitting of the cluster exponential into static and dynamical parts in TCC is similar to that in the mixed-exponentially generated four (MEG4) method of Nakatsuji et al.~\cite{Nakatsuji1985exponentially,Nakatsuji1991mixed}; however, the subsequent formalism is substantially different.

The split cluster operators are explicitly written as,
\begin{equation}
\begin{split}
    \hat{T}^\textrm{active} &= \hat{T}^\textrm{active}_1 + \hat{T}^\textrm{active}_2 \\
    &= \sum_{i,a} t_i^a \hat{a}_a^\dagger \hat{a}_i + \sum_{i,j,a,b} t_{ij}^{ab} \hat{a}_a^\dagger \hat{a}_b^\dagger \hat{a}_i \hat{a}_j,\\
    &\quad i,j,a,b \in \textrm{active space},
\end{split}
\end{equation}
and
\begin{equation}
\begin{split}
    \hat{T}^\textrm{rest} &= \hat{T}^\textrm{rest}_1 + \hat{T}^\textrm{rest}_2 \\
    &=\sum_{i,a} t_i^a \hat{a}_a^\dagger \hat{a}_i + \sum_{i,j,a,b} t_{ij}^{ab} \hat{a}_a^\dagger \hat{a}_b^\dagger \hat{a}_i \hat{a}_j,\\
    &\quad \{i,j,a,b\} \not\subset \textrm{active space},
\end{split}
\end{equation}
where $\hat{a}_a^\dagger$ and $\hat{a}_i$ are the creation and annihilation operators for orbitals $a$ and $i$, respectively, and $t_i^a$, $t_{ij}^{ab}$ are the CC amplitudes.

In conventional CC theory, the amplitudes are obtained by solving the amplitude equations~\cite{Crawford2000introduction}. 
In TCC, the amplitudes within the active space are fixed by CI coefficients via
\begin{align}
\hat{T}_1^\text{{active}} &= \hat{C}_1, \label{eq:t1c1} \\
\hat{T}_2^\text{{active}} &= \hat{C}_2 - \frac{1}{2} \hat{C}^2_1, \label{eq:t2c2}
\end{align}
where $\hat{C_1}$ and $\hat{C_2}$ are the CI singles and doubles operators.
With a constant $\hat{T}^\text{active}$, the remaining amplitudes are solved in the usual CC manner.

In addition to this tailored coupled-cluster singles-and-doubles approach, a perturbative triples (T) correction can be introduced~\cite{Lyakh2011tailored}. This correction is introduced by setting the active-space amplitudes to zero, thereby preventing the double counting of static correlation.

\subsection{Review of quantum selected configuration interaction}

\subsubsection{Basic formulation of QSCI}

The QSCI method~\cite{Kanno2023quantum} samples the quantum state prepared on a quantum computer.
The resulting configurations $\{\ket{\Phi_i}\}$ span a subspace of the Fock space, and an effective Hamiltonian
\begin{equation}
    \hat{H}^\textrm{eff}_{ij} = \braket{\Phi_i | \hat{H} | \Phi_j}
\end{equation}
is constructed and efficiently diagonalized on a classical computer,
\begin{equation}
    \hat{H}^\textrm{eff}\ket{\Psi_\textrm{QSCI}} = E_\textrm{QSCI}\ket{\Psi_\textrm{QSCI}},
\end{equation}
where $E_\textrm{QSCI}$ is the energy of QSCI.
The eigenstate is expressed as
\begin{equation}
    \ket{\Psi_\textrm{QSCI}} = \sum_{i=1}^R c_i\ket{\Phi_i},
\end{equation}
where $c_i$ is the expansion coefficient of the $i$th basis state.
$R$ is the number of retained configurations, which can be truncated based on the sampling frequency. 
Thus, the quantum state is faithfully reconstructed on a classical computer.

\subsubsection{Cartesian product of bitstrings} \label{sec:cartesian}

Finite shot noise can lead to spin-symmetry breaking in the sampled space.
To mitigate this, each sampled determinant can be separated into its $\alpha$- and $\beta$-spin parts,
\begin{equation}
    \ket{\Phi_i} = \ket{\Phi_i^\alpha}\ket{\Phi_i^\beta}, \label{eq:det}
\end{equation}
and the Cartesian product can be formed
\begin{equation} \label{eq:cartesian_product_spin_conf}
    \bigl\{ \ket{\tilde{\Phi}_k} \bigr\} = \bigl\{ \ket{\Phi_i^\alpha} \ket{\Phi_j^\beta}, ^\forall i,j \leq R \bigr\},
\end{equation}
thereby enlarging the subspace spanned by $\{ \ket{\tilde{\Phi}_i} \}$ instead of $\{ \ket{\Phi_i} \}$ and increasing the likelihood that all relevant spin eigenfunctions are included~\cite{Yoshida2025auxiliary,Robledo2024chemistry}.
Although this increases the dimension of $\hat{H}^{\text{eff}}$, it often improves accuracy.

To ensure spin adaptation, we treat the union of the determinants sampled for the $\alpha$- and $\beta$-spin sectors as the configuration set for each spin.
Whenever a determinant of the form Eq.~(\ref{eq:det}) is sampled, the subspace is enlarged by adding its spin-swapped partner $\ket{\Phi_{i'}}$, defined as
\begin{equation}
    \begin{split}
    \ket{\Phi_{i'}^\alpha} &= \ket{\Phi_i^\beta} \\
    \ket{\Phi_{i'}^\beta} &= \ket{\Phi_i^\alpha}. \label{eq:union}
    \end{split}
\end{equation}

\subsection{QSCI-TCC}

\subsubsection{Overview}
\begin{figure*}[htpb]
 \includegraphics[width=\textwidth]{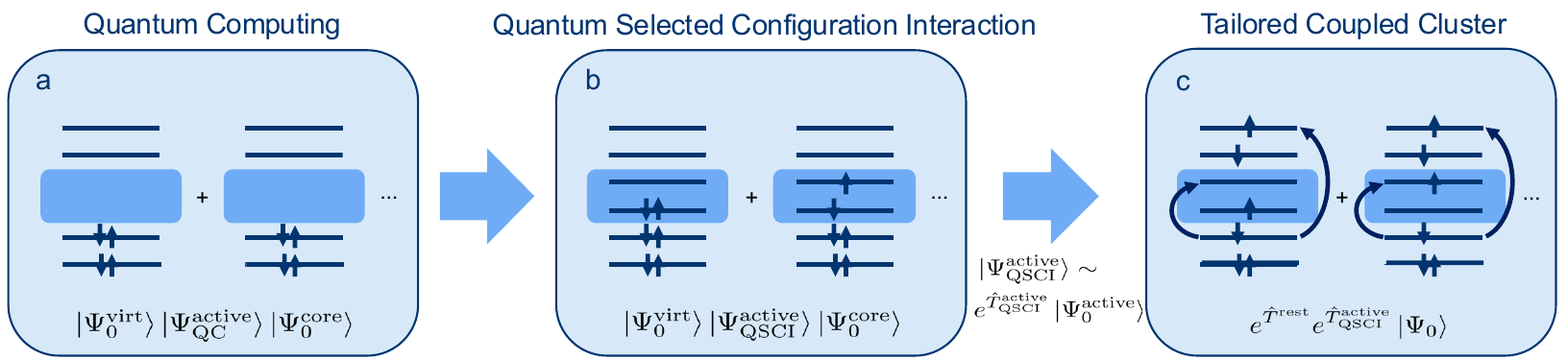}
 \caption{
Schematic depiction of QSCI-TCC. 
(a) A quantum algorithm prepares $\ket{\Psi_\textrm{QC}^\textrm{active}}$. 
(b) QSCI reproduces this state classically as $\ket{\Psi_\textrm{QSCI}^\textrm{active}}$.
(c) TCC then recovers the dynamical correlation.
}
\label{fig:QSCI_TCC_schematic}
\end{figure*}

The proposed QSCI-TCC scheme performs a TCC calculation using a QSCI wave function (Fig.~\ref{fig:QSCI_TCC_schematic}).
First, a quantum algorithm such as VQE or QPE prepares the strongly correlated active-space wave function $\ket{\Psi_\textrm{QC}^\textrm{active}}$ (Fig.~\ref{fig:QSCI_TCC_schematic}~(a)).
Repeated measurements yield bitstrings from which QSCI reconstructs $\ket{\Psi_\textrm{QSCI}^\textrm{active}}$ on a classical computer (Fig.~\ref{fig:QSCI_TCC_schematic}~(b)).
Finally, $\ket{\Psi_\textrm{QSCI}^\textrm{active}}$ is mapped to CC amplitudes, $e^{\hat{T}_\textrm{active}}$ is formed, and the remaining amplitudes are optimized while keeping the active-space amplitudes fixed (Fig.~\ref{fig:QSCI_TCC_schematic}~(c)).

\subsubsection{The (c) correction}

Because TCC at the singles-and-doubles level includes only up to second-order excitations, mapping CI coefficients to CC amplitudes introduces an error in the QSCI-TCC energy $E_{\textrm{QSCI-TCC}}$.
Two main sources contribute to this error: (i) the neglect of triples or higher excitations and (ii) an incomplete set of determinants in QSCI.
To compensate, a corrected energy
\begin{equation} \label{equ_corrected_ernergy}
\begin{split}
    E_{\text{QSCI-TCC(c)}} &= E^{\text{active}}_{\rm \text{QC}} \\
    &+ (E_{\textrm{QSCI-TCC}} - E_{\textrm{QSCI-TCC}}^{\rm \text{active}})
\end{split}
\end{equation}
is introduced~\cite{Erhart2024}.
Here, $E^{\text{active}}_{\rm \text{QC}}$ denotes the total energy obtained within the active-space approximation via quantum computing.
In this study, we substitute the QSCI analog $E^{\text{active}}_{\rm \text{QSCI}}$ for $E^{\text{active}}_{\rm \text{QC}}$.
The last term, 
\begin{equation}
\begin{split}
    E_{\textrm{QSCI-TCC}}^{\text{active}} = \bra{\Psi_0} \hat{H} e^{\hat{T}^\text{active}}\ket{\Psi_{0}},
\end{split}
\end{equation}
corresponds to the energy obtained immediately after tailoring, i.e., before the amplitudes outside the active space are optimized.

\section{Computational details} \label{sec:comput_details}

This section summarizes the settings used for the numerical evaluation of QSCI-TCC.

We used PySCF 2.2.1~\cite{PySCF,PySCF2} for quantum chemical calculations, including Hartree--Fock (HF), CCSD, CCSD(T), and full configuration interaction (FCI).
The basis set used for each calculation is specified individually in the corresponding subsection of Sec.~\ref{sec:results}. 

Simulations of quantum computing were performed with Quri-Parts 0.20.3.
VQE simulation runs relied on Chemqulacs~\cite{chemqulacs} and used two ansatzes: the disentangled unitary coupled cluster singles and doubles (UCCSD) ansatz~\cite{kutzelniggQuantumChemistryFock1982,*kutzelniggQuantumChemistryFock1983,*kutzelniggQuantumChemistryFock1985,*bartlettAlternativeCoupledclusterAnsatze1989,*kutzelniggErrorAnalysisImprovements1991,*taubeNewPerspectivesUnitary2006,*evangelista2019exact,Peruzzo2014variational} and the \texttt{GateFabric} ansatz~\cite{Anselmetti2021Local}.
Fermionic operators were mapped to qubits using the Jordan--Wigner transformation.

We performed the $10^7$ shot measurements for sampling quantum states in the investigation of energy curves in Sec.~\ref{sec:bond_dissoc}.
The influence of shot count is analyzed in detail in Sec.~\ref{sec:meas}.
Effective Hamiltonians generated by QSCI were diagonalized using the \texttt{kernel\_fixed\_space} function in PySCF.

\section{Results and discussion} \label{sec:results}

\subsection{Bond dissociation \label{sec:bond_dissoc}}

We benchmark QSCI-TCC on two prototypical strong-correlation problems: the simultaneous dissociation of both O--H bonds in H$_2$O and the triple bond dissociation in N$_2$.
These systems are widely used to test methods designed for MR situations.

\subsubsection{Simultaneous dissociation of the OH bonds in H\textsubscript{2}O}

\begin{figure*}
    \centering
    \includegraphics[width=\textwidth]{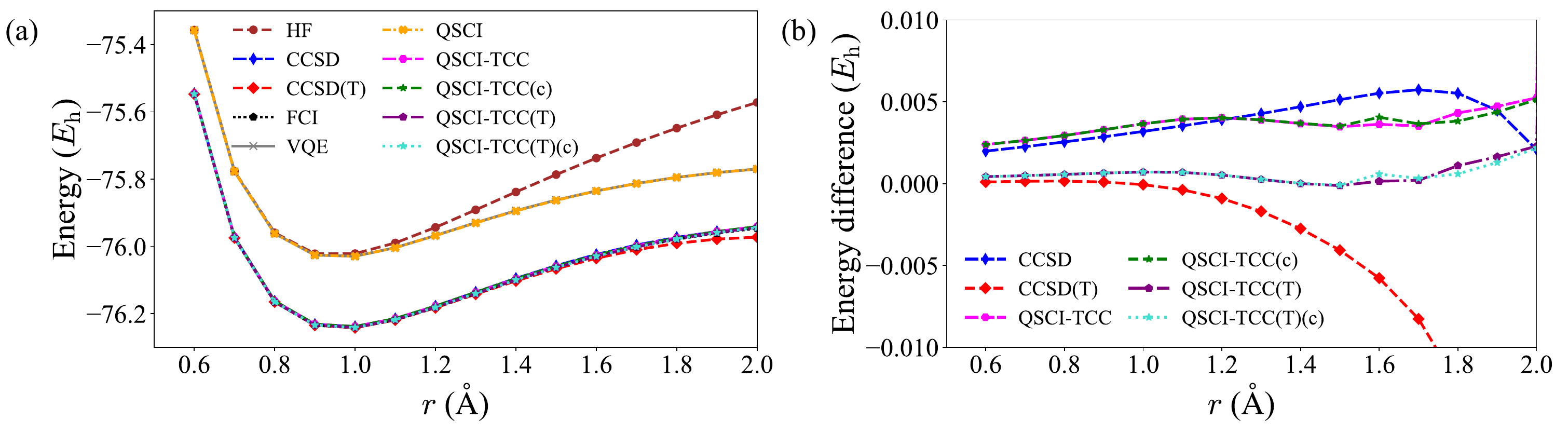}
    \caption{(a) Energy curve and (b) energy difference compared to FCI for simultaneous O--H bond dissociation in H$_2$O at the cc-pVDZ basis set level. The active space is (8e, 6o), and the ansatz for VQE is the UCCSD ansatz.}
    \label{fig:H2O}
\end{figure*}

Figure~\ref{fig:H2O}~(a) shows the potential energy curves for the simultaneous stretching of the two O--H bonds.
Starting from the HF baseline, static correlation within the (8e, 6o) active space is introduced and captured.
Both active-space QSCI and VQE lower the energy relative to HF, and the resulting gap widens as the bond lengths $r$ become larger.

The subsequent TCC calculations incorporate dynamical correlation from the remaining orbitals, further lowering the energy across the entire region.

We also plot CCSD, CCSD(T), and exact FCI energies for comparison.
Their curves are nearly indistinguishable from the QSCI-TCC curves; however, CCSD(T) begins to depart from the others at approximately $r=2.0$ \AA.

Fig.~\ref{fig:H2O}~(b) shows the deviation from FCI. 
Although CCSD(T) begins to diverge from the FCI reference at approximately $r=2.0$ \AA, the QSCI-TCC(T) error remains almost flat over the entire range. This indicates that the static correlation captured in the QSCI wave function stabilizes the description even at large bond separations.  
The difference between the CCSD and corresponding QSCI-TCC energy error curves highlights the effect of amplitude embedding; both are accurate for this system.
The additional (c) correction alters the energies only marginally and is therefore negligible here.

\subsubsection{Triple-bond dissociation in N\textsubscript{2}}

\begin{figure*}
    \centering
    \includegraphics[width=\textwidth]{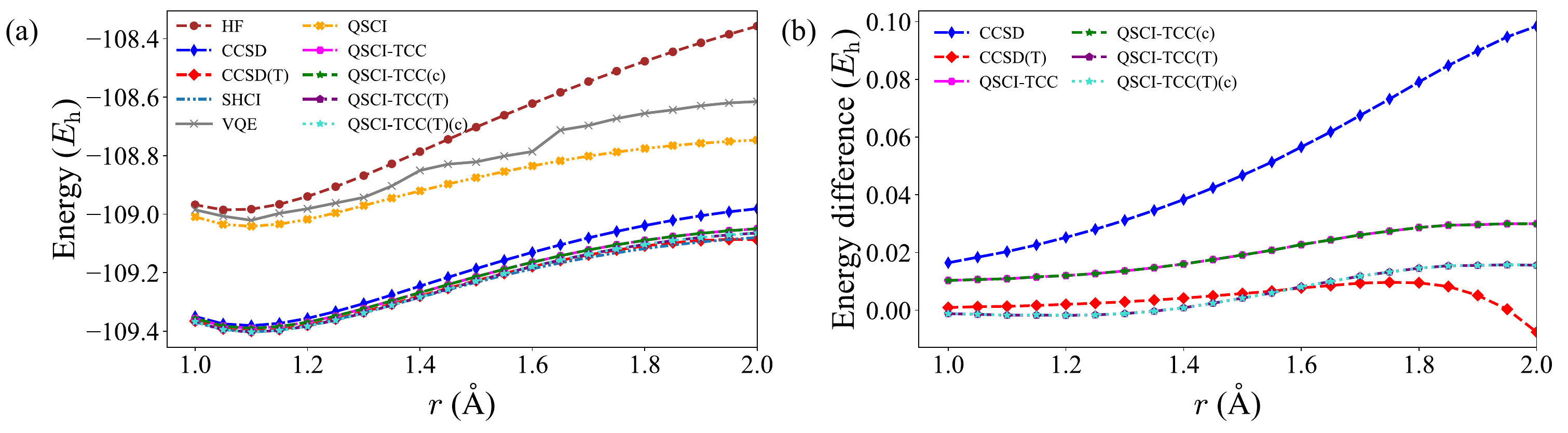}
    \caption{(a) Energy curve and (b) energy difference compared to SHCI for N$_2$ dissociation at the cc-pVTZ basis set level. The active space is (6e, 6o). The \texttt{GateFabric} ansatz with three layers is used for VQE.}
    \label{fig:N2}
\end{figure*}

Fig.~\ref{fig:N2}~(a) presents the potential energy curves for the dissociation of the triple bond in N$_2$.
Similar to H$_2$O, static correlation recovered within the (6e, 6o) active space is essential for a qualitatively correct description of the dissociation region.
Our VQE calculation, which employs a three-layer \texttt{GateFabric} ansatz, fails to produce a smooth energy curve; nevertheless, QSCI delivers a stable result even when the underlying VQE wave function is imperfect.
This robustness can be attributed to the fact that the effective Hamiltonian is built from computational basis states that belong to the exact wave function, regardless of the quality of the trial state.
All methods that include correlation over the full set of orbitals---such as QSCI-TCC and its variants, CCSD(T), and the semistochastic heat-bath CI (SHCI) reference---converge to an essentially common curve on the scale of Fig.~\ref{fig:N2}~(a). The only exception is that CCSD underestimates the energy. The referential SHCI data are taken from Ref.~\cite{Lee2022Data}.

For further evaluation of accuracy, Fig.~\ref{fig:N2}~(b) shows the energy deviation from SHCI.
The CCSD error grows monotonically over the scanned range.
In the region from 1.8 to 2.0 \AA, the CCSD(T) error also starts to increase, reflecting the well-known breakdown of CCSD(T) for N$_2$ dissociation.
By contrast, the QSCI-TCC and QSCI-TCC(T) error curves remain nearly flat, mirroring the behavior already observed for H$_2$O.
This stability indicates that the CC amplitudes not only describe dynamical correlation but also successfully incorporate static correlation by embedding the QSCI wave function.

Similar to the case of water, the additional (c) correction has a negligible effect on the resulting energy curves.

\subsection{Dependence on measurement shots \label{sec:meas}}

We investigated how the number of measurement shots influences the statistical uncertainty of QSCI-TCC(c).
The QSCI sampling was varied as $10^3$, $10^4$, $10^5$, and $10^6$ shots, and these measurements were repeated $10^3$ times independently.
All simulations employed a two-layer \texttt{GateFabric} ansatz and (6e, 6o) active space, corresponding to twelve Jordan--Wigner qubits.
Two N$_2$ geometries were considered: $r=1.1$ \r{A}, close to the equilibrium bond length, and $r=2.2$ \r{A}, far into the strongly correlated regime.
Identical configuration sets obtained from repeated measurements were first grouped. Then, a single TCC calculation was performed for each unique set while we separately recorded how many times that set appeared. 
To restore the spin symmetry, the data were analyzed with and without taking the union of the separated $\alpha$- and $\beta$-spin determinants, as shown in Eq.~(\ref{eq:union}).

\begin{figure*}[htpb]
 \includegraphics[width=0.9\textwidth]{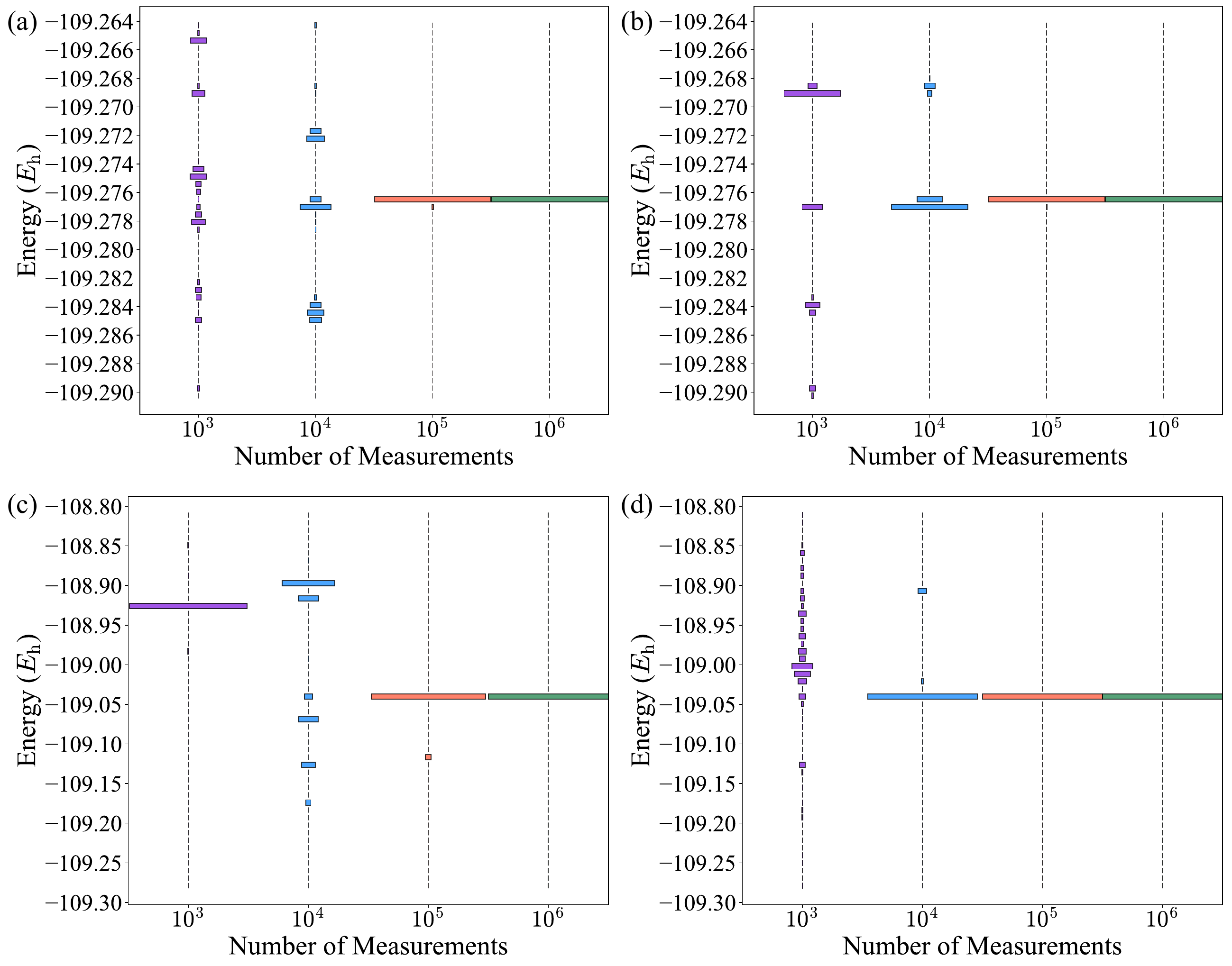}
 \caption{Influence of the number of measurement shots on QSCI-TCC(c) for N$_2$ at the cc-pVTZ basis-set level. 
 (a) $r=1.1$ \r{A} with the union option;
 (b) $r= 1.1$ \r{A} without the union; 
 (c) $r= 2.2$ \r{A} with the union;
 (d) $r= 2.2$ \r{A} without the union.}
\label{fig:meas}
\end{figure*}

Fig.~\ref{fig:meas} summarizes the results.
In every panel, the QSCI-TCC(c) energy converges rapidly as the shot count increases.
We observed that the $1.0 \times 10^5$ shots are sufficient in each condition, but the convergence behaviors differ between the two bond lengths. The details are described below.

At the near-equilibrium geometry ($r=1.1$~\AA; Figs.~\ref{fig:meas}~(a)~and~(b)), the repeated $10^5$ shot measurements sample 22 grouped subsets with the union option and 41 subsets without the union.
The corresponding QSCI-TCC(c) energies fall into a single bin of width $5.3 \times 10^{-4}~E_\mathrm{h}$.
Because the molecule is close to equilibrium, a relatively small subset of dominant determinants, such as the HF state, is sufficient.
Hence, once the limited set is included, the energy is considered to be determined with high precision.

In the strongly correlated regime ($r=2.2$~\AA; Figs.~\ref{fig:meas}~(c)~and~(d)), $10^5$ shots yield only two distinct determinant sets with the union option and one without it.
Consequently, each peak in the $10^5$-shot histograms of panels~(c) and~(d) corresponds to the energy of a single unique subset.
The pronounced peak near $-109.04~E_\mathrm{h}$, for example, arises from a subset containing $361 (=19^2)$ configurations.
Here, the Cartesian product enlargement allows for a more diverse set of basis states, making $10^5$ shots sufficient.

In the cases of less than $10^5$ shots, the observed energy distribution is discrete rather than continuous, reflecting the binary occurrence of key determinants in the sampled set.
The energies can be lower than the converged value because CC theory is not variational in nature.

From the above observations, we conclude that the $10^5$-shot measurements achieve an error below 1~kcal/mol ($\sim 1.6 \times 10^{-3}~E_\mathrm{h}$) in this setting.
No apparent difference is observed between the union and non-union treatments in this study; in both cases, the number of measurements required for convergence remains of the same order.

The present requirement of $10^5$ shots outperforms earlier quantum-classical hybrid TCC studies.
Computational basis tomography required $3 \times 10^7$ shots to achieve the standard deviation of $0.61 \times 10^{-3}$~$E_\text{h}$ in the $r=5.355$ Bohr ($\sim 2.8$~\AA) case at the cc-pVDZ level~\cite{Erhart2024}. 
Although this previous study was conducted under a challenging bond-distance setting, the relative phases need to be determined for computational basis tomography~\cite{Kohda2022quantum}; therefore, QSCI is considered superior.
In addition, at the cc-pVDZ level, the estimated result of the matchgate shadows tomography demonstrates that $1.0 \times 10^5$ and $2.2 \times 10^6$ shots were required to obtain chemical precision at $r=1.1$ and $2.2$~\AA, respectively~\cite{Scheurer2024}, where the latter bond length is more strongly correlated. These comparisons underscore the shot efficiency of QSCI-TCC.

\section{Conclusions} \label{sec:conclusions}

We introduced QSCI-TCC, a quantum-classical hybrid method that integrates the QSCI wave function into the TCC framework.
QSCI provides a scalable, shot-efficient way to reconstruct a quantum-prepared state on a classical computer. 
The wave function reconstructed by QSCI contains no statistical error, only model error, providing a stable state for classical post-processing.
Embedding the resulting CI coefficients as fixed active space amplitudes, the subsequent CC calculation supplies the missing dynamical correlation, yielding a balanced treatment of static (strong) and dynamical (weak) correlation effects.

Benchmark calculations for simultaneous O--H bond dissociation in H$_2$O and triple-bond dissociation in N$_2$ show that, even in regions of the dissociation curves where CCSD or CCSD(T) begins to deteriorate, QSCI-TCC and QSCI-TCC(T) continue to deliver accurate results. 
Thus, the static correlation captured by quantum computation, although a simulator has been used in this study, is successfully embedded within the coupled cluster description.

Shot-count investigations on N$_2$ (6e, 6o) active space further show that QSCI-TCC(c) converges rapidly with respect to the number of shots.
We found that $10^5$-shot measurements are sufficient to reach chemical precision.
This represents an order-of-magnitude improvement in measurement efficiency compared to an earlier implementation using matchgate shadows tomography in the $r=2.2$~\AA~case, where it is a strongly correlated scenario.

Efficient quantum state tomography---using QSCI as well as other tomography methods---remains a key research topic. 
Most recently, Lenihan et al. have introduced a related study on another shadows-based protocol combined with a surrogate CC model~\cite{Lenihan2025excitation}.

It is worth mentioning a critical issue with QSCI and the proposed improvements in recent years.
In strongly correlated systems, the number of Slater determinants with non-trivial coefficients can grow exponentially with system size.
To mitigate this, enhanced sampling schemes that utilize real-time evolution have been proposed~\cite{Sugisaki2024Hamiltonian,Mikkelsen2024quantum,Yu2025quantum}.
Besides, Reinholdt et al. have compared QSCI with a conventional selected CI intensively~\cite{Reinholdt2025critical}.
Improving sampling efficiency, therefore, remains an important avenue for future work.

%\vspace{4ex}
\section*{Ackowledgements}
The authors are grateful to Takuma Murokoshi for his technical support.
This project was supported by funding from the MEXT Quantum Leap Flagship Program (MEXTQLEAP) through Grant No. JPMXS0120319794.
This study was conducted as part of a joint research in the Quantum Software Research Hub (Grant No. JPMJPF2014), and further supported by TOPPAN Digital Inc.

% The \nocite command causes all entries in a bibliography to be printed out
% whether or not they are actually referenced in the text. This is appropriate
% for the sample file to show the different styles of references, but authors
% most likely will not want to use it.
%\nocite{*}

\bibliography{apssamp}% Produces the bibliography via BibTeX.

\end{document}